\documentclass[conference]{IEEEtran}
\IEEEoverridecommandlockouts
\usepackage{cite}
\usepackage{amsmath,amssymb,amsfonts}
\usepackage{algorithmic}
\usepackage{graphicx}
\usepackage{textcomp}
\usepackage{xcolor}
\usepackage{graphicx}
\usepackage{hyperref}
\usepackage[frozencache=true,cachedir=.]{minted}
\usepackage{pgfplots}
\pgfplotsset{compat=1.16}
\usepackage{balance}
\usepackage{adjustbox}
\def\BibTeX{{\rm B\kern-.05em{\sc i\kern-.025em b}\kern-.08em
    T\kern-.1667em\lower.7ex\hbox{E}\kern-.125emX}}
\begin{document}

\title{ValueNet: A Natural Language-to-SQL System \\ that Learns from Database Information}

%\author{\IEEEauthorblockN{1\textsuperscript{st} Ursin Brunner}
\author{\IEEEauthorblockN{Ursin Brunner}
\IEEEauthorblockA{%\textit{dept. name of organization (of Aff.)} \\
\textit{ZHAW Zurich University of Applied Sciences}\\
Winterthur, Switzerland\\
Ursin.Brunner@zhaw.ch}

\and
%\IEEEauthorblockN{2\textsuperscript{nd} Kurt Stockinger}
\IEEEauthorblockN{Kurt Stockinger}
\IEEEauthorblockA{%\textit{ZHAW Datalab} \\
\textit{ZHAW Zurich University of Applied Sciences}\\
Winterthur, Switzerland \\
Kurt.Stockinger@zhaw.ch}
}

\maketitle

\begin{abstract}
Building natural language (NL) interfaces for databases has been a long-standing challenge for several decades. The major advantage of these so-called NL-to-SQL systems is that end-users can query complex databases without the need to know SQL or the underlying database schema. Due to significant advancements in machine learning, the recent focus of research has been on neural networks to tackle this challenge on complex datasets like Spider. Several recent NL-to-SQL systems achieve promising results on this dataset. However, none of the published systems, that provide either the source code or executable binaries, extract and incorporate values from the user questions for generating SQL statements. Thus, the practical use of these systems in a real-world scenario has not been sufficiently demonstrated yet.

In this paper we propose \emph{ValueNet light} and \emph{ValueNet } -- two end-to-end NL-to-SQL systems that incorporate values using the challenging Spider dataset. The main idea of our approach is to use not only metadata information from the underlying database but also \emph{information on the base data} as input for our neural network architecture. In particular, we propose a novel architecture sketch to extract values from a user question and come up with possible value candidates which are not explicitly mentioned in the question. We then use a neural model based on an encoder-decoder architecture to synthesize the SQL query. Finally, we evaluate our model on the Spider challenge using the \emph{Execution Accuracy} metric, a more difficult metric than used by most participants of the challenge.
Our experimental evaluation demonstrates that \emph{ValueNet light} and \emph{ValueNet} reach state-of-the-art results of  67\% and 62\% accuracy, respectively, for translating from NL to SQL whilst incorporating values.
\end{abstract}

\begin{IEEEkeywords}
NL-to-SQL, natural language interface, neural networks, transformers
\end{IEEEkeywords}

\section{Introduction}
\label{sec:introduction}

Building NL-to-SQL systems is a long-standing challenge in both the database and the natural language processing community. Being able to query databases and other structured data in natural language gives users without knowledge in a query language access to a large amount of information. Therefore, a natural language interface has often been regarded as the most powerful database
interface\cite{NLIPowerfull}.

\begin{figure}[ht]
\centering
\includegraphics[width=3.5in,clip,keepaspectratio]{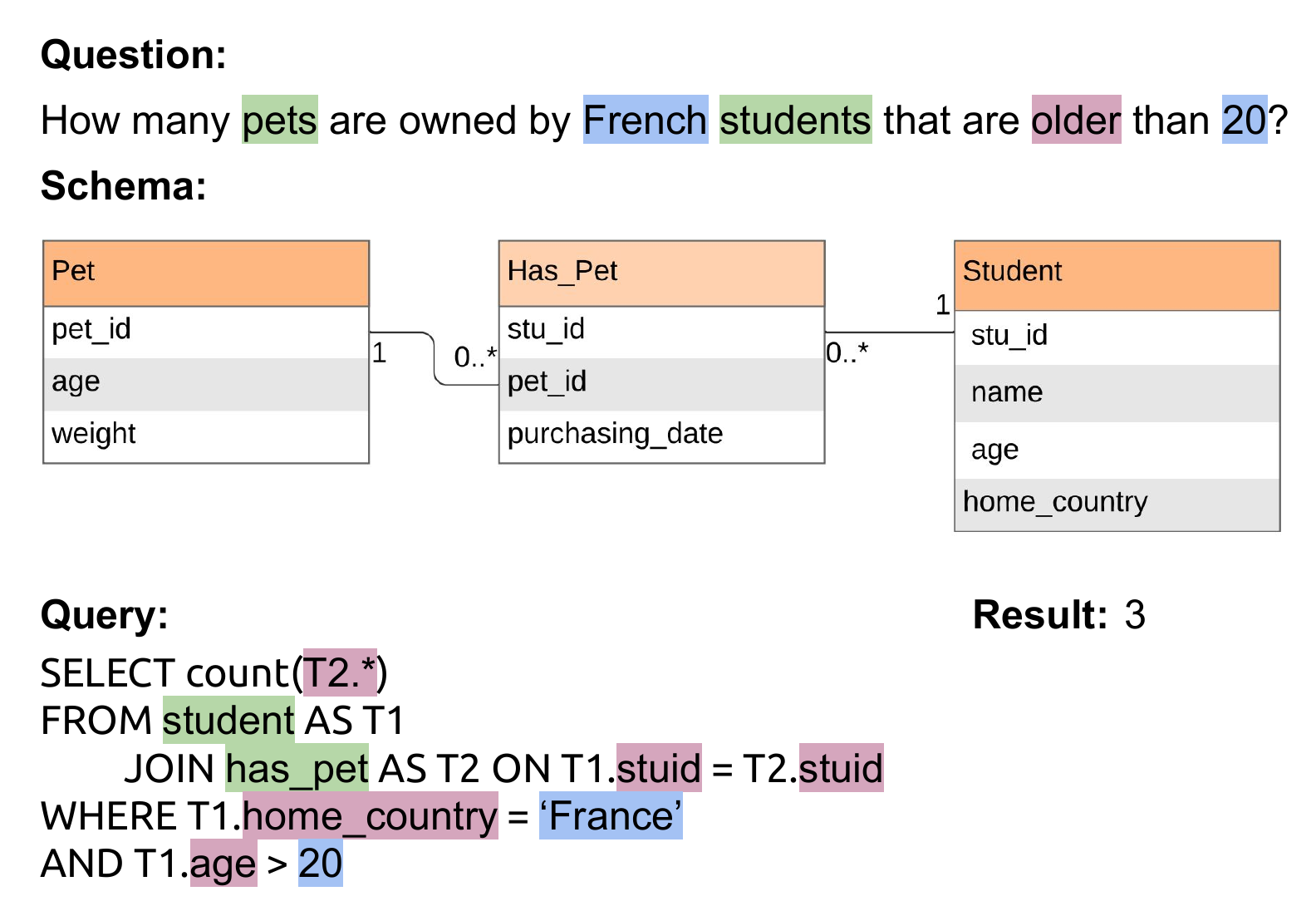}
\caption{A typical NL-to-SQL system synthesizes and executes a SQL query given a natural language question and a database schema. Besides synthesizing a valid SQL sketch (non-highlighted words), the system has to select the correct tables \emph{(green)}, columns \emph{(red)} and values \emph{(blue)}}.
\label{figure:initial_example}
\end{figure}

At the same time proprietary systems such as Google's \emph{Assistant} or Amazon's \emph{Alexa} are improving the way users can access large knowledge bases with natural language. It is therefore an integral focus of any open data effort to offer users a similarly convenient interface to query data by natural language.

In Figure \ref{figure:initial_example} we see the typical NL-to-SQL flow: Given a natural language question and a database schema, the system has to synthesize a valid SQL query. Once executed, this query should return the answer the user was looking for. Let us consider the query {\it How many pets are owned by French students that are older than 20?}. This example shows some of the challenges that such a system is confronted with. The fact that "older" is most probably referring to the column \emph{age} (in the \emph{Student} table rather than in the \emph{Pet} table!) cannot be directly extracted from the question. The same principle applies to the fact that "French students" most probably refers to an entry of the column  \emph{home\_country} containing the {\it value} "France". In other words, the token "French" of the natural language question does not directly map to a value in the {\it base data} of the underlying database. The latter challenge is the focus of our paper, i.e. how to build an end-to-end neural architecture that {\it takes values into account}. 

The goal is to generate complex, nested SPJ\footnote{Select-Project-Join} SQL statements including WHERE-clauses. The WHERE-clauses are constructed by predicting the {\it correct table, column and base data value} (e.g. student.home\_country = 'France') that correspond to the respective value in the natural language question (e.g. "French students"). We will see below that taking values into account has so far received little attention from the current state-of-the-art of neural network-based NL-to-SQL systems. 

%Over the last years, a large body of research focused on new NL-to-SQL architectures \cite{affolter2019comparative}. With the introduction of large-scale datasets like \emph{WikiSQL}\cite{WikiSQL}, advanced neural approaches started to become popular. These approaches often used an encoder-decoder architecture to train a neural network end-to-end on a large amount of question/query pairs.

%\emph{WikiSQL} comprises 80,000 question/query pairs and enables the use of such neural architectures. However,  \emph{WikiSQL} lacks the complexity of real world scenarios. With only one table per database (and therefore no \mintinline[breaklines]{SQL}{JOIN}s), as well as the lack of advanced operators like \mintinline[breaklines]{SQL}{ORDER BY}, \mintinline[breaklines]{SQL}{GROUP BY} or nested queries, simple sketch-based decoders such as e.g. SQLova\cite{SQLova} started soon to reach accuracy scores of 90\% and more on the test set.

%The \emph{Spider}\cite{Spider} challenge released by the end of 2018 poses a new interesting challenge in the field of NL-to-SQL. Not only does  \emph{Spider} overcome the shortcomings of \emph{WikiSQL} with respect to complex queries, it also requires the participants to implement a NL-to-SQL system handling unseen databases. Providing a total of 200 databases over 138 domains, the participants have to avoid overfitting on specific domains or known database schemas in order to generalize well on unseen questions and database schemas.

The release of \emph{Spider} dataset motivated several research groups to provide contributions in 2019/2020 and has become the de-facto standard for evaluating NL-to-SQL systems. While classical neural architectures\cite{DongLapata,SQLNet} achieved rather low accuracy scores on the \emph{Spider} leaderboard\footnote{https://yale-lily.github.io/spider (last accessed on Oct 8, 2020)} (4.8\%, 12.4\%), more advanced approaches like \emph{IRNet}\cite{IRNet} and \emph{RAT-SQL}\cite{RatSQL} score significantly higher with accuracies over 60\%.

The \emph{Spider} challenge was released with two evaluation metrics and leaderboards:
\begin{itemize}
\item \textbf{Exact Matching Accuracy}: The metric measures if a predicted query is equivalent to the gold query. Thanks to its smart \emph{component matching}, this metric is tolerant towards ordering issues (e.g. \mintinline[breaklines]{SQL}{SELECT A, B} vs \mintinline[breaklines]{SQL}{SELECT B, A}). Unfortunately though the metric does not evaluate {\it values} such as "French" or "20" in Figure \ref{figure:initial_example} -- which is essential in real-world scenarios. In other words, this metric only considers the schema of the database and not the base data. The reason is that queries do not need to be executed against a database for this type of evaluation.
\item \textbf{Execution Accuracy}: Measures if the results of both predicted and gold query are the same by {\it executing} them against a real database. To pass this evaluation, a NL-to-SQL system not only has to predict the correct SQL sketch and select the right tables/columns, but has also to identify the correct {\it values} and place them in the right order. 
\end{itemize}

As of today, most \emph{Spider} contributions focus on the \emph{Exact Matching Accuracy} evaluation (see "Leaderboard - Exact Set Match without Values"). While a good score on this leaderboard is definitely an achievement (it requires a system to predict a complex SQL-sketch and select the right tables and columns), it is still a simplification of a fully fledged NL-to-SQL system. It does not address the challenging task of generating and selecting {\it values} and it still abstracts from several other difficulties addressed throughout this paper. Although there are a few contributions on the Spider leaderboard that evaluate their systems based on the more complex {\em Execution Accuracy} (see "Leaderboard - Execution with Values"$^2$), there is currently {\it no published paper or published source code available}. Hence, there is currently no description and systematic evaluation of a neural network-based system for the Spider dataset that translates natural language to SQL and takes values into account.

In this paper we make the following {\bf contributions}:
\begin{itemize}

\item To the best of our knowledge, we provide the first {\it detailed discussion as well as the source code of an NL-to-SQL system} to synthesize a full SQL query including values on the challenging Spider dataset using {\em Execution Accuracy}. Our approach is thus a step forward in building an end-to-end-system to generate complex, nested SPJ-queries that are typical of real-world scenarios. By providing the source code, our {\it approach is also reproducible} - which is often difficult to achieve when systems do neither provide the source code nor executable binaries. We achieve a state-of-the-art accuracy of up to 67\% on the Spider dataset using the more challenging \emph{Execution Accuracy} metric. 

\item We provide {\it two novel NL-to-SQL architectures}: (1) \emph{ValueNet light} - A system which selects the correct values from a given list of possible ground truth values and then synthesizes a full query including the chosen values. (2) \emph{ValueNet} - An end-to-end architecture which extracts and generates value candidates given only natural language questions and the database content. \emph{ValueNet} then uses these value candidates to synthesize a full query including values equivalent to \emph{ValueNet light}.

\item We show that the difference in performance between \emph{ValueNet} and \emph{ValueNet light} is relatively small given a {\it strong generative approach for the candidate generation}. This indicates that if we come up with the right value candidates, the neural model is capable of selecting them correctly.
\end{itemize}

The paper is organized as follows. In Section 2 we define the problem of NL-to-SQL in more detail. In Section 3 we in describe the high-level architecture of our NL-to-SQL system. In Section 4 we discuss the detailed architecture with focus on ValueNet light and ValueNet. Our experimental results are given in Section 5. In Section 6 we review the related work. Finally, we draw conclusions in Section 7.

\section{Problem Definition}

\subsection{Intermediate Representation via Abstract Syntax Tree}

Before we introduce our end-to-end architecture for translating from natural language to SQL, we describe the concept of {\it Abstract Syntax Trees} (AST). The idea is to use ASTs as an intermediate representation in the overall translation process - rather than translating directly to SQL. The advantage of using an AST as an immediate representation is to overcome the so-called \emph{mismatch} problem ~\cite{IRNet}, where end-users rarely pose a question as detailed as necessary in order to directly synthesize a SQL query. Therefore, abstracting the details of SQL enables the system to understand the question/SQL query pairs more reliably as shown previously \cite{guo-etal-2019-towards, cheng2019learning}. Another advantage is that an intermediate representation enables the system to be independent of a specific query language.

For our approach we do not create a grammar for such a representation from scratch, but extend the context-free grammar \emph{SemQL} of \emph{IRnet}\cite{IRNet}. We call our extended version \emph{SemQL 2.0}. 

\begin{figure}[h!]
\centering
\includegraphics[width=3.5in,clip,keepaspectratio]{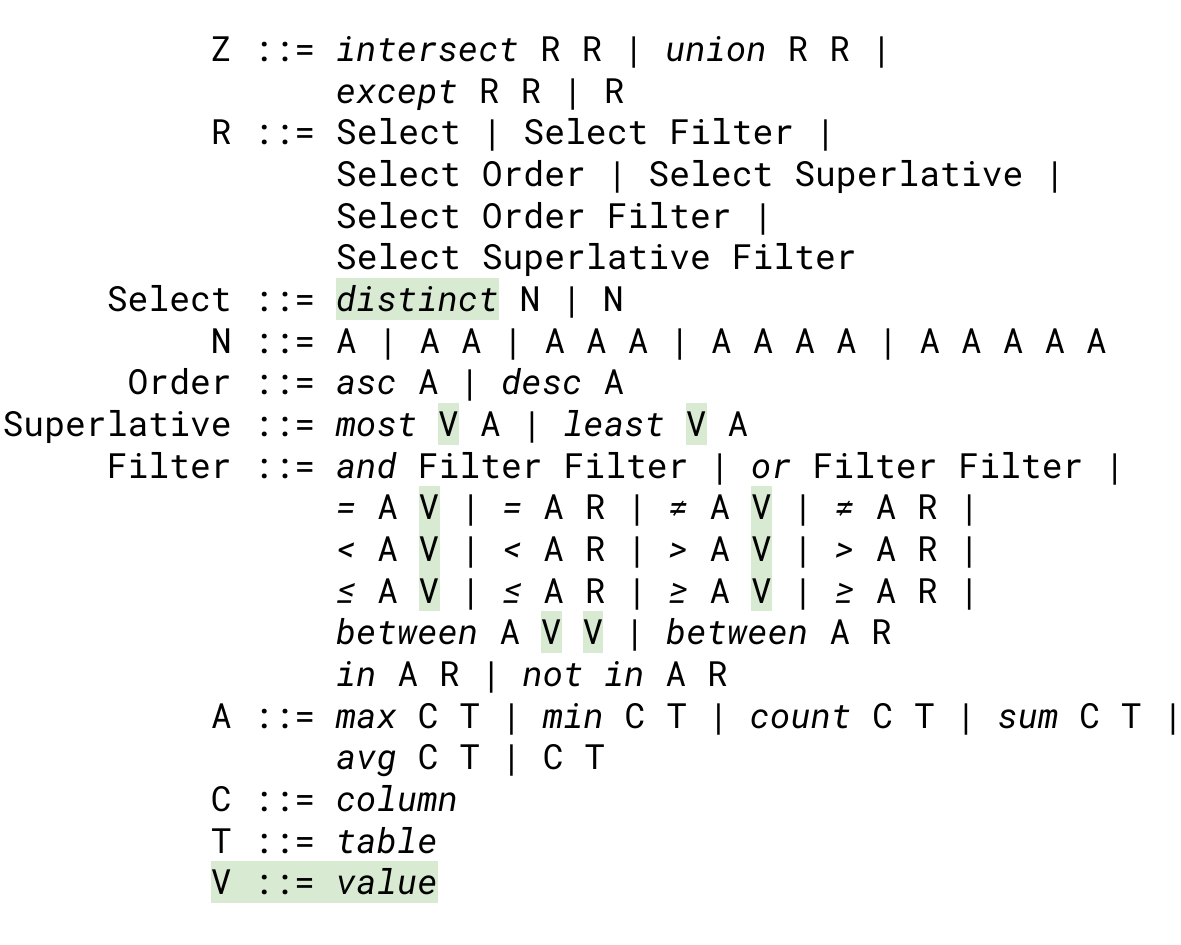}
\caption{SemQL 2.0 Grammar. Our contributions to the previously published SemQL 1.0 grammar are highlighted in green.  The identifiers to the left of the \emph{::=} represent all possible actions in the grammar. Italic tokens represent SQL operators or references to tables/columns/values. The pipe separates different implementations of an action.}
\label{figure:semql_2_0_grammar}
\end{figure}

Figure \ref{figure:semql_2_0_grammar} shows the complete grammar for SemQL 2.0 that can handle the major relational operators such as {\it select, project, join, union, intersect} as well as {\it aggregations and complex, nested queries}. We extend the grammar introduced in \emph{IRnet}\cite{IRNet} mainly by the value representation \emph{V} - highlighted in the figure in green.

\subsection{Problem Statements: NL to AST to SQL} 

Given the grammar of SemQL 2.0,  we will now describe the problem statements for translating a natural language question (NLQ) to SQL using a supervised machine learning approach such as neural networks. In particular, we can distinguish between two types of sub problems: (a) NLQs that require synthesizing a SQL-statement by predicting only information that is contained in the {\it database schema}, i.e. tables and columns. (b) NLQs that require synthesizing a SQL-statement by predicting also values that are contained in the {\it base data}. In summary, the set of options for prediction in problem (b) is much larger than in problem (a).

%Problem (a) has been tackled by previous work based on the Spider dataset, while to the best of our knowledge, we are the first ones to tackle the harder problem (b) on the Spider dataset.

%Predicting values is distinguished from the prediction of SQL sketch/tables/columns by the fact that a neural network model can not simply choose from a fixed set of options (candidate values).

Let us define the term \emph{set of options} for predicting matching tables, columns and values using our running example query {\it How many pets are owned by French students that are older than 20?}. The set of options for predicting \emph{tables} corresponds to the total number of tables contained in the database schema. For instance, in our example shown in Figure \ref{figure:initial_example}, we have three tables. The set of options for predicting \emph{columns} is the total number columns of the database schema -- which is 10 in our example. However, the set of options for predicting \emph{values} cannot just be looked up in the database schema. Here, we potentially require an inverted index over each column of the entire database. Since the number of values is far greater than the number of tables and columns, predicting the right value is a computationally more complex problem than predicting the right table or column.

%To understand this difference we will first analyze how columns, tables and the SQL-sketch are predicted. Afterwards we describe how predicting values deviates from this approach. 

\subsubsection{{\bf Predicting Columns, Tables and SQL-Sketches}}
\label{subsec:column_tables_sqlsketch}
For predicting columns and tables, the set of options is given by the database schema. In principle, one can simply provide all table names and column names as input to a neural network such that it can learn the mapping between the question tokens and the database schema. As an example, have a look again at Figure \ref{figure:initial_example}. During the training phase, the neural network learns that the question tokens \emph{"older than"}  should be mapped to column \emph{age} rather than e.g. to \emph{stu\_id}. %This knowledge is incorporated into the column encoding of \emph{age} and is used by the pointer network of the decoder to choose the correct column.

%It is important to realize that the encoder does not have to learn such attention from scratch. Thanks to the encoder model being heavily pre-trained on a large text corpus (see BERT\cite{BERT} or RoBERTa\cite{RoBERTa} for details on the pre-training), it already learned that there is a stronger correlation between \emph{older than} and \emph{age} than e.g. \emph{older than} and \emph{stu\_id}.

After predicting tables and columns, the next step is to predict a SQL-sketch which works slightly differently than predicting tables and columns. Although the neural network chooses from a finite set of options, this set changes dynamically. As an example, assume that the last chosen action in the SQL-sketch is an \emph{Order} action. By definition of the SemQL 2.0 grammar (see Figure \ref{figure:semql_2_0_grammar}), the only possible options now are action \emph{asc A} and \emph{desc A}. The neural network can therefore only choose from options given by the SemQL 2.0 grammar. These options dynamically change depending on the preceding node in the SemQL 2.0 tree.

%\newpage

\subsubsection{{\bf Predicting Values}}
\label{subsec:values}

\begin{figure}[ht]
\centering
\includegraphics[width=3.5in,clip,keepaspectratio]{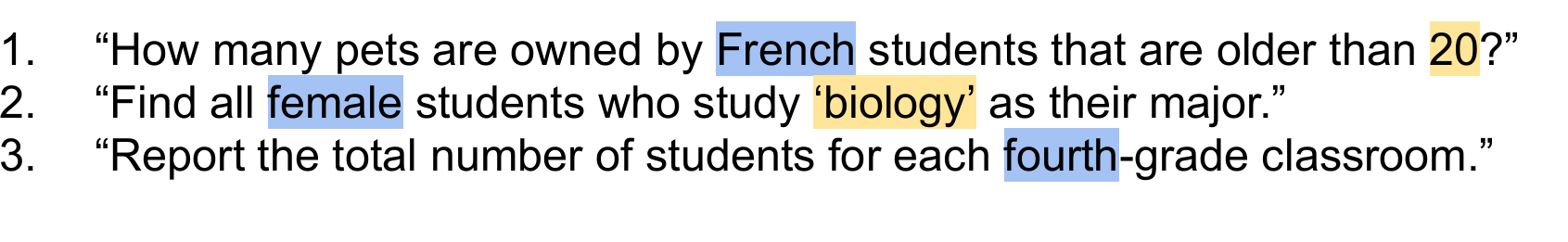}
\caption{Three examples of values (blue) which are typically not directly derivable from the text. 
%For instance, the tokens \emph{"French students"} might not be directly contained in the database but refer the to students with home country \emph{France}. %
The yellow values, on the contrary, are directly derivable from the text.}
\label{figure:values_not_included_in_question}
\end{figure}

In contrast to predicting columns, tables, and SQL-sketches, no fixed set of options exists for values. While one could assume that all possible values are contained in the question, this is not always the case. In Figure \ref{figure:values_not_included_in_question} we see three examples of values which are not typically directly derivable from the text. The terms \emph{French students} are most probably referring to a student with the home country \emph{France}. The term \emph{female} might refer to students whose gender is equal to \emph{'F'} and the fourth-grade classroom might refer to a table \emph{classroom} with grade \emph{4}.

\begin{figure}[ht]
\centering
\includegraphics[width=3.5in,clip,keepaspectratio]{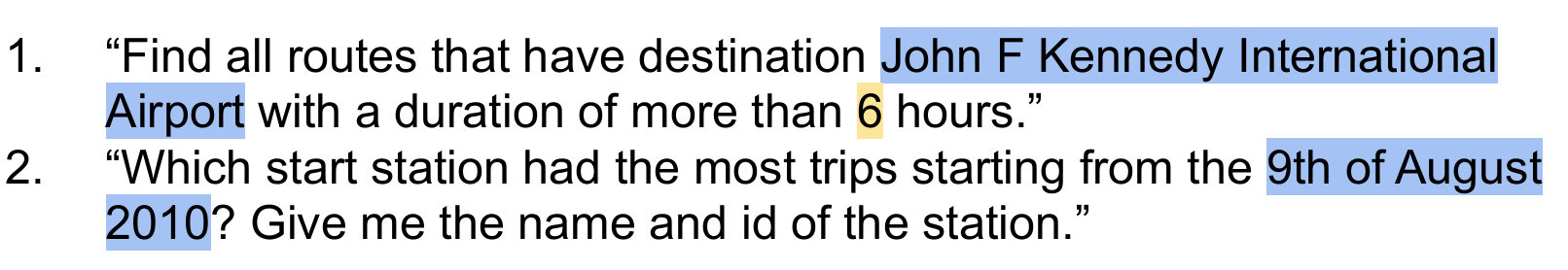}
\caption{Examples of values which result in a large number of possible value candidates.}
\label{figure:values_large_combinatorics}
\end{figure}

Even in cases where all information is available in the question, there is often a large number of possible candidates for a given value. In Figure \ref{figure:values_large_combinatorics} we see two such examples where a natural language question can have a large number of possible {\it value candidates} contained in the base data. While \emph{John F Kennedy International Airport} definitely refers to a column containing airports, we have no idea how this value is stored in the database. It could be anything from \emph{JFK} over \emph{John F Kennedy} to the full term \emph{John F Kennedy International Airport}. Similarly, we do not know how the date in the second example, i.e. \emph{9th of August 2010}, needs to be formatted to match the given value in the database, which could be stored as "2010-08-09".

The challenge in all these examples can be reduced to one single problem: unlike columns, tables, and SQL sketches, we do not know the set of options for a given value.

\section{ValueNet: High Level Architecture}
\label{sec:architecture}

In this section we describe the high level architecture of ValueNet to translate a natural language question to SQL. 

\begin{figure}[ht]
\centering
\includegraphics[width=3.5in,clip,keepaspectratio]{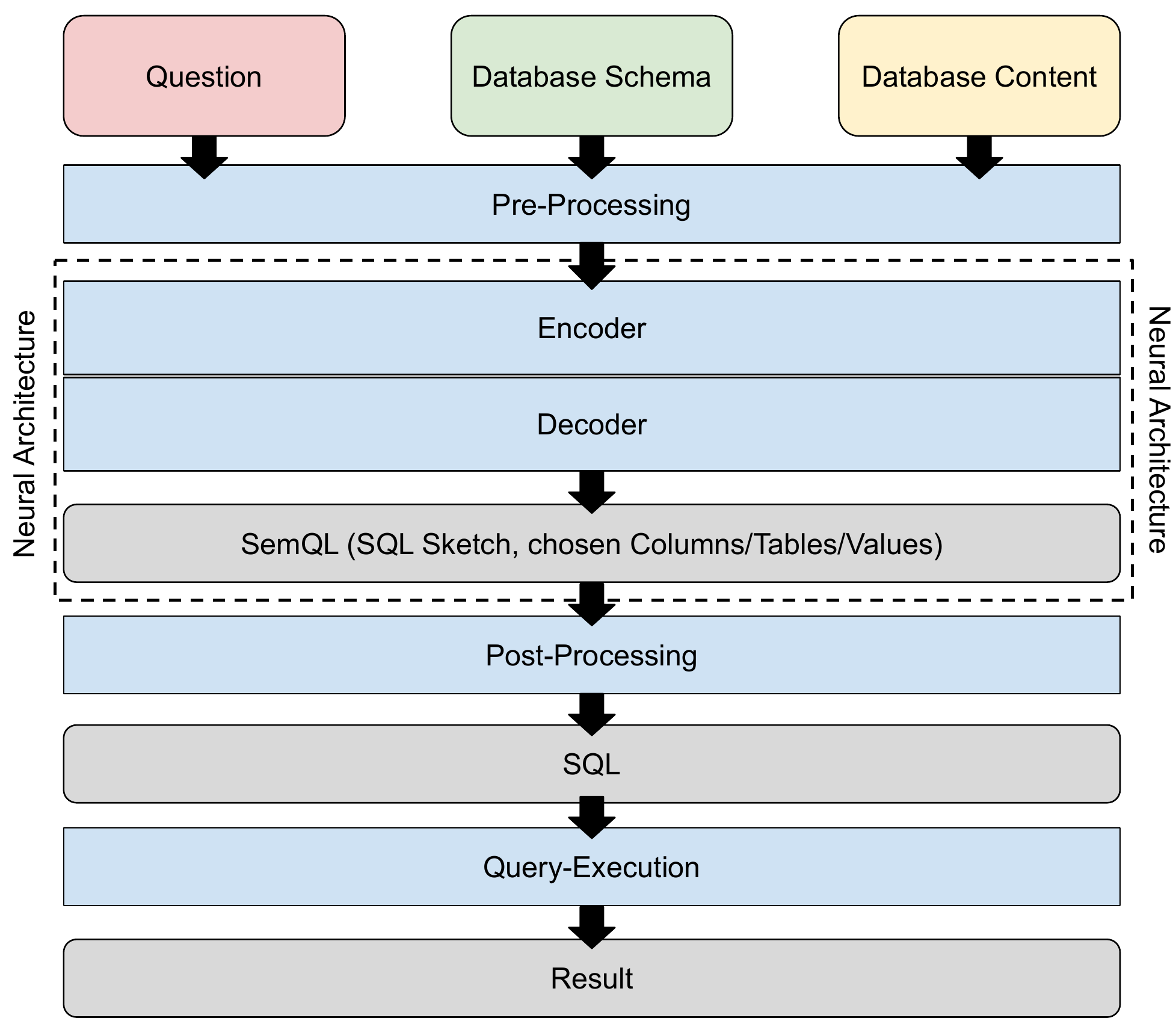}
\caption{ValueNet Architecture Overview.}
\label{figure:architecture_overview}
\end{figure}

Figure \ref{figure:architecture_overview} shows the end-to-end process of our NL-to-SQL system. As input our system expects a question in natural language, the schema of the database, and access to the content of the database, e.g. via an inverted index as used in \cite{blunschi2012soda}. The output is a SQL statement, that once executed, returns the data the user asked for. 

We will elaborate the steps of our architecture based on the initial example in Figure \ref{figure:initial_example} with the question \emph{How many pets are owned by French students that are older than 20?} 

Note that previously published systems that were evaluated against the Spider dataset did not incorporate {\it values} such as "French" or "20". However, these systems would simply fill in a placeholder value (e.g. \emph{'1'}) for each value, as the \emph{Exact Matching Accuracy} of Spider does not validate values. %where no access to the base data is required and hence the problem is much simpler to solve.

\subsection{Pre-Processing}
\label{subsec:preproessing}
During pre-processing we try to achieve two tasks: The first task is to come up with good \emph{value candidates}, which we describe in detail in Section \ref{sec:ValueNet}. The second task is to come up with useful \emph{hints} for predicting tables, columns and values. These hints are made available to the neural network, i.e. the subsequent components of our architecture, as an additional source of information besides the natural language question and the database schema. The intuition of this process is to give the neural network model "hints" which are easy to establish (e.g. by looking at the database content) and support it to take the correct decision when predicting SQL columns, tables, SQL-sketches and values.

%By using pre-trained transformer models, the classical NLP pre-processing steps like stop-word removal or stemming get obsolete. The transformer-specific pre-processing with word pieces and word embeddings we discuss in Section \ref{subsec:NeuralArchitecture}.

\subsubsection{{\bf Question Hints}}
\label{subsubsec:QuestionHints}
\begin{figure}[ht]
\centering
\includegraphics[width=3.5in,clip,keepaspectratio]{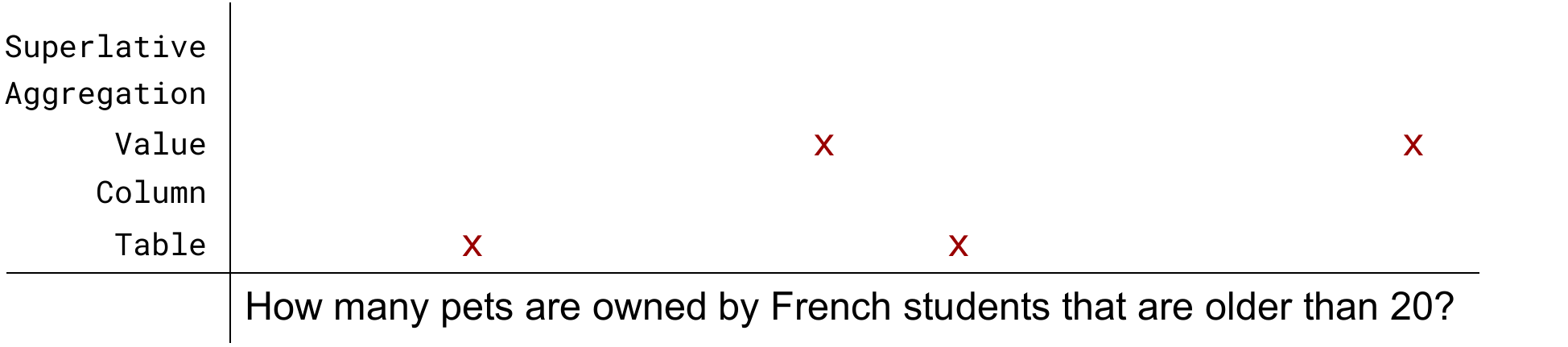}
\caption{Classify question tokens by finding matches in the schema (for columns/tables) or database content (for values).}
\label{figure:PreProcessingQuestion}
\end{figure}

For each token in the question we try to figure out if it refers to a \emph{table}, a \emph{column}, a \emph{value}, an \emph{aggregation} or a \emph{superlative}. See Figure \ref{figure:PreProcessingQuestion} for an example on how we classify the tokens of a question. The intuition behind this is that those tokens are probably the most important ones when synthesizing the query. 

For now we do not use any advanced NLP methods to find matches between question tokens and schema information - we simply apply stemming to all words and look for exact matches. This process can be improved as part of future work by using word embeddings and other advanced techniques. Keep in mind though, that this pre-processing is just a simple way to provide {\it prior knowledge} to the neural model. More complex relationships, like e.g. the fact that the token \emph{older} in Figure \ref{figure:PreProcessingQuestion} refers most probably to the column \emph{age} in table \emph{Student}, are established by the encoder part of our neural network.

\subsubsection{{\bf Schema Hints}}
\label{subsubsec:SchemaHints}

\begin{figure}[ht]
\centering
\includegraphics[width=3.5in,clip,keepaspectratio]{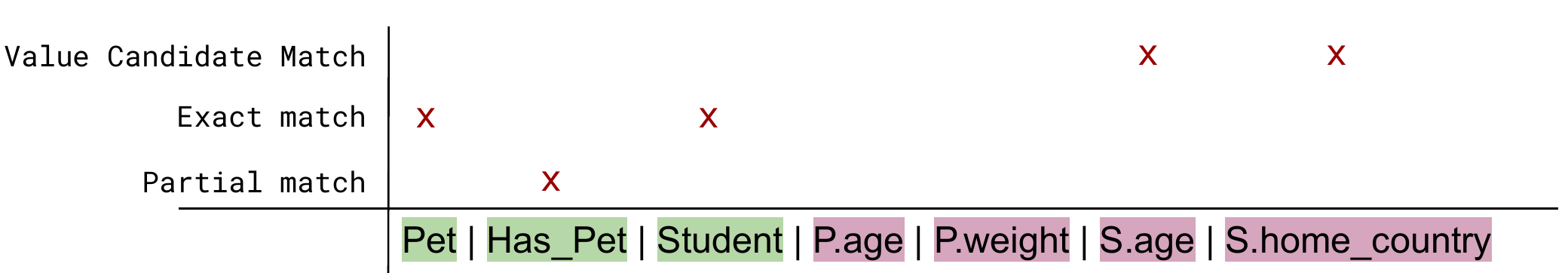}
\caption{Schema Hints, based on matches between tables/columns and the tokens in the question. This example refers to a subset of the schema in Figure \ref{figure:initial_example}.}
\label{figure:PreProcessingSchema}
\end{figure}

Schema hints are basically the inverse version of the \emph{Question Hints}. We want to know if a column or table has been mentioned in the natural language question. Again, the intuition is to give hints to the neural network about the importance of a column/table. See Figure \ref{figure:PreProcessingSchema} for an example. If a table or a column matches exactly with a token in the question (e.g. in Figure \ref{figure:initial_example} for \emph{"pet"} and \emph{"student"}) we classify it as an \emph{exact match}. If it is only a partial match (e.g. for the token \emph{"pet"} and table \emph{Has\_Pet}), we classify it as such. A special case is the class \emph{value candidate match}: We apply this class if a value candidate has been found in the database in a specific column. As an example take the value token \emph{"20"} of our running example. As we found it in column \emph{Student.age}, we classify this column as a \emph{value candidate match}.

\subsection{Neural Architecture}
\label{subsec:NeuralArchitecture}

The next two process steps, {\it Encoder} and {\it Decoder}, are what we call the \emph{Neural Architecture}. This component is the core of our NL-to-SQL system. As input it receives the \emph{question}, \emph{schema} and \emph{hints} from the pre-processing step and synthesizes a query represented in \emph{SemQL 2.0}.

In particular, the encoder tries to {\it encode information about the question and the schema} into a low dimensional space. Based on the encoded information, the decoder then tries to {\it synthesize a valid query} step by step.
Applied to our example, we want the Neural Architecture to learn that e.g. \emph{"How many"} should be translated to \mintinline[breaklines]{SQL}{SELECT count()} and that the word \emph{"older"} should refer to the column \emph{Student.age}.

%The intuition here is that an advanced neural model can learn to translate complex natural language questions to SemQL. The assumption is further that such a neural network based on a natural language architecture is more advanced in understanding the nuances of language than, for example, a template-based approach and is therefore superior on this task.  

%The neural model consists of an Encoder-Decoder architecture which is common for natural language processing tasks of all kinds. While the encoder focuses on encoding information about the source (in our case the natural language question, the database schema and the value candidates) into a lower dimensional space, the decoder focuses on generating the output representation. In our case the generated output representation is the query in form of an abstract syntax tree based on the SemQL 2.0 grammar.

\subsubsection{{\bf Encoder}}
\label{subsubsec:encoder}

Our encoder is based on a {\it pre-trained transformer architecture}~\cite{AttentionIsAllYouNeed} which is used in most recent NL-to-SQL systems ~\cite{SQLova, IRNet, RatSQL}. Transformer architectures have been used for different tasks such as natural language translation~\cite{AttentionIsAllYouNeed}, natural language generation~\cite{GPT} and recently also for entity matching as part of data integration~\cite{brunner2020entity}. The intuition is that such attention-only architectures of transformers generate better representations of natural language sequences than classical recurrent neural networks (RNNs). Hence, transformer architectures often yield better results on many natural language processing tasks than conventional neural networks. 

%Furthermore, due to the missing recurrency of transformers, they are easier to parallelize on modern hardware and therefore more efficient in training. By pre-training such large transformer models on huge amounts of text data in an unsupervised manner~\cite{BERT, RoBERTa}, transformer models became even more powerful. IRNet\cite{IRNet} showed that simply replacing the encoder with BERT\cite{BERT} leads to a performance increase of about 7-8\% on the Spider challenge.

%\begin{figure*}[h]
%\centering
%\includegraphics[width=\textwidth,clip,keepaspectratio]{resources/Question_Schema_Encoding.pdf}
%\caption{Encoding of natural language question and database schema tokens. The differences to the IRNet encoder are highlighted. Each color visualizes a  token type: light gray for question tokens, light green for table tokens, red for column tokens and dark green for special tokens such as separators. The black connectors on top visualize the attention built up during encoding between question- and schema tokens, e.g. between question token "older" and column token "age". The yellow boxes represent word embeddings. }
%\label{figure:QuestionSchemaEncoding}
%\end{figure*}

%A critical task when using transformer encoders is to correctly encode the input. See Figure \ref{figure:QuestionSchemaEncoding} for an example. The input in our case is the question (already tokenized), the schema (table- and column names already tokenized). 

Our encoder is an extension of IRNet's encoder. The main difference is that our encoder receives not only information about the database schema but also {\it value candidates}, extracted from the database content. Hence, our encoder can also learn correlations between tokens of an NL question and the actual values in the database.
%We further sub-tokenize all input with the  WordPiece\cite{WordPiece} segmentation algorithm to solve the out-of-vocabulary problem. We then choose for each sub token the correct word embedding, which we input into the encoder.

%One important aspect to notice in Figure \ref{figure:QuestionSchemaEncoding} is the large green encoding \emph{Enc T$_1$}. Such a multi-token encoding gets created if a table or column name consists of more than one token as e.g. column \emph{home\_country} in the initial example (Figure \ref{figure:initial_example}). The decoder though has to choose full columns and tables, e.g. \emph{home\_country} and not single tokens (\emph{home} or \emph{country}) of a column/table. We therefore have to build a single representation from such multi-token columns/tables. We do this by using a bi-directional LSTM into which we sequentially feed all tokens of a column/table. We then keep the final hidden state as encoding representing that column/table.

\subsubsection{{\bf Decoder}}
\label{subsubsec:decoder}

The decoder receives the question/table/column/value-encodings from the encoder as input and outputs a query represented as an AST. The decoder consists of an LSTM architecture \cite{hochreiter1997long} in combination with multiple Pointer Networks\cite{PointerNetwork} to choose tables, columns and values.

\subsection{Post-Processing}
\label{subsec:PostProcessing}
The post-processing step is mainly a deterministic SemQL2.0-to-SQL transformation. We also have to incorporate the selected values, which we will explain in more detail in Section \ref{sec:ValueNet}.

\subsubsection{{\bf Translating SemQL2.0 to SQL}}
Transforming SemQL2.0 to SQL is done by traversing the SemQL2.0 syntax tree from its root to leaf nodes. Most actions of the SemQL2.0 grammar can be transformed directly to their SQL equivalent. %An \emph{Order} action, for example, is transformed simply to an \mintinline{SQL}{ORDER BY} SQL clause. The more abstract \emph{Superlative} action is transformed into \mintinline{SQL}{ORDER BY Col1 DESC LIMIT n}.
%The \emph{Filter} action is slightly more complicated. If there is no aggregation involved, the filter is transformed into a \mintinline{SQL}{WHERE} clause (nested if necessary). If the filter is applied to an aggregated column, it is transformed into a \mintinline{SQL}{HAVING} clause.
%As a rule of thumb, an aggregation causes all other columns in the \emph{Select} to be included in a \mintinline{SQL}{GROUP BY} clause.

\subsubsection{{\bf Handling Relationships}}
\label{subsubsec:Relationships}
Note that the original SemQL does not know about relationships but only about tables used in \emph{Filter, Select, Order} and \emph{Superlative} actions. The reason for this design choice is the fact that users most likely do not mention all the required tables in their questions. As an example take the database schema in Figure \ref{figure:initial_example}. When a user poses the question \emph{"Give me all student names with pets older than 14 years"}, she mentions the tables \emph{Pet} and \emph{Student} as she needs them for \emph{Filtering} and \emph{Selection}. She will though most likely not mention the bridge table \emph{Has\_Pet}. %In a large real world database schema there are often multiple tables between the tables a user mentions.

However, a proper SQL query needs to join all mentioned tables by using the bridge tables. A common approach is to transform the database schema into an undirected graph, where the vertexes are tables and edges are primary-key/foreign-key relationships. One can then build up all \mintinline{SQL}{JOIN}s deterministically by finding the shortest path between two known vertices (tables) by e.g. using the Dijkstra algorithm. As soon as we deal with more than two tables, an approximation algorithm for the \emph{Steiner tree}\cite{AlgSteinerTree} problem will solve the problem of connecting all $N$ vertices by the shortest path even more elegantly.

Unfortunately, we found this approach to be too simplistic when we switched to the Spider \emph{Execution Accuracy} metric. Note that the \emph{Exact Matching Accuracy} does not validate which columns are used to join two tables. It is enough to correctly predict the tables of the join without specifying the \mintinline{SQL}{ON} clause (e.g. \mintinline{SQL}{A INNER JOIN B} instead of \mintinline{SQL}{A INNER JOIN B ON A.A = B.A}), which is the approach used by \emph{IRNet}. 

Obviously this does not hold true anymore when executing queries using the \emph{Execution Accuracy} metric as we do. Here, leaving out the join restriction results in a cross join yielding a Cartesian product of all rows. We therefore extended the schema graph by incorporating the primary/foreign key columns for each relationship edge.  

%\begin{figure}[h]
%\centering
%\includegraphics[width=3.5in,clip,keepaspectratio]{resources/Schema_Graph_Flight.pdf}
%\caption{A simplified version of the Spider database \emph{flight\_2} visualized as a graph.}
%\label{figure:SchemaGraphFlight}
%\end{figure}

%It does not hold as soon as there are multiple relationships between two tables as we do not know which one to choose then. As an example see Figure \ref{figure:SchemaGraphFlight}. When building up the join between \emph{Flights} and \emph{Airports} we have to know if we need to join via the \emph{SourceAirport} or \emph{DestAirport} foreign key. We can therefore not infer the correct JOIN without extracting more information. Tackling this challenge, is part of our future work.

%In large real world databases with tables far away from each other in the schema graph there is a further challenge: there might exist multiple possible paths between two tables and it is not a given that the shortest one is always the one the user is looking for.

\section{ValueNet - Detailed Architecture}
\label{sec:ValueNet}
In this section we introduce \emph{ValueNet light} and \emph{ValueNet} -- two novel NL-to-SQL systems which learn from database information by considering both the database schema and the base data. While \emph{ValueNet light} assumes that a set of value options is provided, \emph{ValueNet} builds up a set of options by itself. The main contributions of \emph{ValueNet} and \emph{ValueNet light} are a {\it novel pre-processing approach} and an {\it enhanced encoding} step (see architecture in Figure \ref{figure:architecture_overview}).

\subsection{ValueNet light}
\label{subsec:ValueNetLight}
Let us assume the true values in the database in the first sample sentence of Figure \ref{figure:values_large_combinatorics} are \emph{'JFK'} and \emph{'6'}. If we had an oracle that provided us a set of options with these values, our neural model would only have to pick the right value at the right time when synthesizing a query. This would work because the encoder most probably establishes more attention between \emph{John F Kennedy International Airport} (the tokens in the natural language question) and \emph{JFK} (the value option from the oracle) rather than between \emph{John F Kennedy International Airport} and \emph{'6'}. %The system could therefore choose the right value at the right time during query synthesizing in the decoder (see Section \ref{subsubsec:decoder}).

This is what we do with \emph{ValueNet light}. We assume that all values of a query have been established upfront and are now available as a set of options. How to establish the values is not part of ValueNet light. One possible way to accomplish this, as the authors of Spider suggest\cite{SpiderConversational}, would be to interact with the user in a question-answer conversation flow in order to establish all needed information for a query.

\begin{figure*}[ht]
\centering
\includegraphics[width=\textwidth,clip,keepaspectratio]{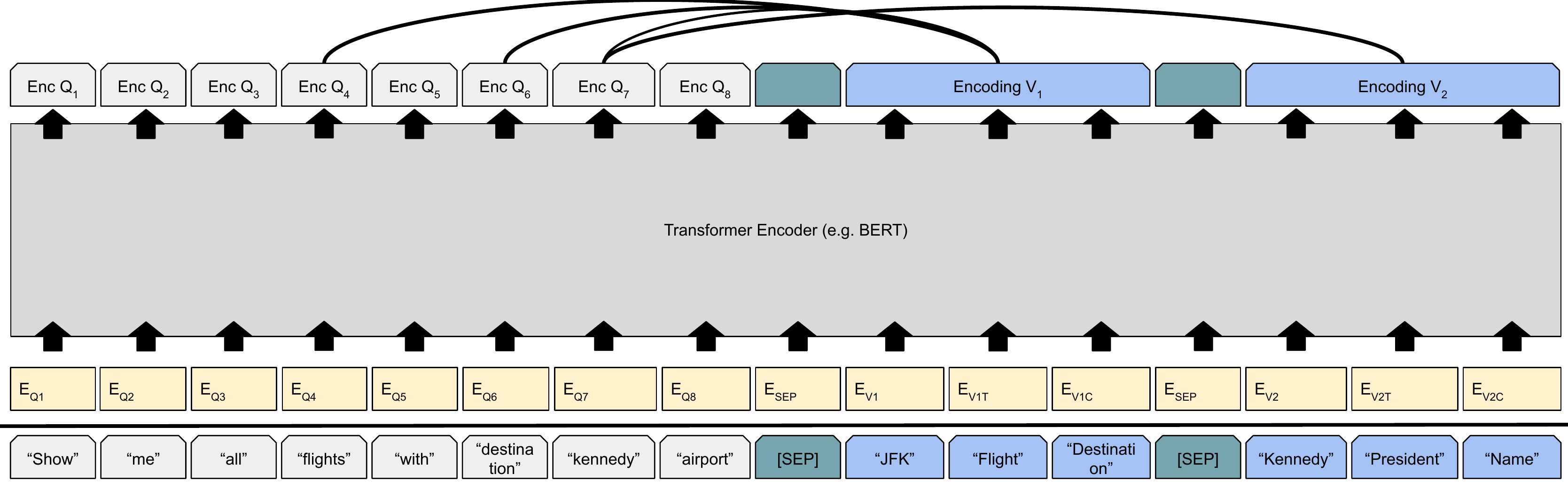}
\caption{Encoding of value candidates and question. The black connectors on top visualize the attention built up during encoding between question tokens and the actual values in the database. The large blue value encodings at the output of the encoder visualize the summarizing of a value together with its location (in a certain table column) by an LSTM. The encodings of columns and tables have been omitted for better readability.} 
\label{figure:value_candidates_encoding}
\end{figure*}

For all experiments with \emph{ValueNet light} we compile the set of value options from the ground truth for each given example. This approach complies with the \emph{Execution Accuracy} metric of Spider.
\emph{ValueNet light} then encodes all these values as part of the input as visible in Figure \ref{figure:value_candidates_encoding} (blue value encodings). For instance, the (blue) values "JFK", "Flight", and "Destination" have established a strong connection (attention) with the question tokens "flights", "destinations" and "kennedy".

In the next step, the neural model selects the right value encodings while synthesizing a query.

% In the same way as for column- and table names we use an LSTM on top of the encodings to summarize multi-token values to one single encoding. Consider, for example, the values \emph{'John F Kennedy International Airport'} and \emph{'6'}, this is intuitively understandable: The decoder has to decide between these two values, there is no possibility it could only go for the token \emph{'Airport'} as this value not exists.

%\subsubsection{Post-processing}
In the deterministic post-processing step we format the value given the predicted data type of the column. If the column is, for example, of the type text, we add quotes to it. If it is of the type integer, we make sure a floating point is not provided. In the case that the SQL sketch predicts a \emph{Filter} action of type \emph{like}, we further extend the value with the SQL wildcard character $\%$.

\subsection{ValueNet}
\label{subsec:ValueNet}
In contrast to \emph{ValueNet light} we propose with \emph{ValueNet} an end-to-end NL-to-SQL system which solves the harder problem where no value candidates are given upfront. To come up with value candidates we propose an architecture sketch consisting of the following steps:
\begin{itemize}
\item {\it Value Extraction}: Extracting values from the question by using named entity recognition (NER) and heuristics. 
\item {\it Value Candidate Generation:} Generating value candidates by searching similar values in the database and by manipulating the extracted values (e.g. n-grams). 
\item {\it Value Candidate Validation:}  Reducing the set of value candidates by looking them up in the database. 
\item {\it Value Candidate Encoding:} Encoding the remaining candidates together with information about the tables and columns they have been found in. This is then the input for the neural architecture. 
\item {\it Neural Architecture:} Continuing similar to \emph{ValueNet light} based on the architecture described in Section \ref{sec:architecture}.
\end{itemize}

We will now explain these steps in more detail.

\subsubsection{{\bf Value Extraction}}
Given a natural language question, we use two different named entity recognition (NER) models to extract potential values. As a first approach, we implement a custom NER model based on a transformer\cite{AttentionIsAllYouNeed} architecture leveraging the popular Transformers\cite{TransformerLibrary} library. As a second approach we use a commercial NER API\footnote{ \href{https://cloud.google.com/natural-language/docs/analyzing-entities}{https://cloud.google.com/natural-language/docs/analyzing-entities}}.

While a custom NER model has the advantage of being able to fine tune on specific domains or datasets, this approach poses the danger of overfitting on certain types of values. Using a commercial NER API reduces this risk, though obviously at the danger of worse results, as it is not specifically trained to the task at hand. Alternatives to this NER API are other popular off-the-shelf libraries for NER as e.g. the SpaCy\cite{spacy2} \emph{EntityRecognizer}.

%A good NER model does not only detect and classify values, but also classifies its sub-groups. A date value, for example, can be extracted in \emph{year}, \emph{month} and \emph{day} separately. This simplifies \emph{Candidate Generation} in the next step. 

In addition to a stochastic NER model we suggest {\it deterministic heuristics} to extract some types of values. We use the following three simple heuristics to identify candidate values: (1) Content in quotes: e.g. \emph{Whose head's name has the substring 'Ha'?}. (2) Capitalized terms: e.g. \emph{Show all flight numbers with aircraft Airbus A340-300.} (3) Single letters: e.g. \emph{When is the hire date for those employees whose first name does not contain the letter M?}

While a custom NER model can easily be trained to detect these types of values, heuristics allow for augmenting the results of an off-the-shelf solution.

%\newpage

\subsubsection{{\bf Value Candidate Generation}}
\label{subsubsec:CandidateGeneration}
After extracting values from the question, we need to generate value candidates. While for numeric values the extracted value itself is most likely the only necessary candidate, the process of value candidate generation is essential for all {\it text-based value types}. For ValueNet we implemented three simple methods of value candidate generation -- one based on \emph{string similarity}, one based on \emph{handcrafted heuristics} and one based on {\it n-grams}.

Generating value candidates through {\it similarity} to existing values in the database is trivial in theory but challenging to implement efficiently. To measure similarity between a text value extracted from the question and values in the database, one can use either classical text distances\cite{EditDistance} or distances based on word embeddings\cite{Word2Vec}. One then only has to scan the database for values with a similarity above a certain threshold.

The need for an efficient implementation stems from the fact that this pre-processing step has to be executed at inference time of each question and its complexity is bound to the size of the database. As the users are, at that point, actively waiting for an answer, the generation of candidates should ideally take less than a second. By using smart indexes and computationally cheap methods for blocking/indexing\cite{Christen}, this effort can be optimized. We further use the Damerau–Levenshtein\cite{DamerauLevenshtein} distance to measure similarity between tokens because of its good trade-off between accuracy and run time.

%As all the databases of the Spider set are rather small (the largest database, \emph{baseball\_1} has \textless  50 tables and around 300,000 entries per table at most) there is no need for advanced methods. However, we use proper parallelization to search for similar values as we plan to apply our method also to larger databases. 

A second way to generate value candidates is through {\it handcrafted heuristics}. This is necessary due to the fact that databases have a specific (but reoccurring) approach to implement certain data types. We currently use the following heuristics: (1) Classic \emph{gender} values, for example, are often implemented as a VARCHAR(1) column with content \emph{'F'} or \emph{'M'}. (2) {\it Boolean} data types are often implemented by a numeric column with value 0 and 1. (3) {\it Ordinals} as e.g. in \emph{"...fourth-grade students..."} are usually implemented as an integer column. (4) Months (e.g. \emph{August}) are often part of a full date column. By using a wildcard (e.g. \emph{8/\%}) once can find them.

While such simple handcrafted heuristics do not generalize to every database, they are a good starting point to bootstrap a generative model which learns such patterns in a more dynamic way.

A third approach for value candidate generation is to use {\it n-grams}. We use this technique for all extracted values with more than one token. For example, a value like \emph{'Kennedy International Airport}' generates one trigram, two bigrams, and three single words as value candidates.

\subsubsection{{\bf Value Candidate Validation}}

Depending on the (1) similarity thresholds, the (2) number of values extracted from the natural language question, and the (3) total number of values in the database, candidate generation might result in a large set of potential candidates. As we see in the experiments for \emph{ValueNet light} and \emph{ValueNet} in Section \ref{sec:experiments}, the number of candidates has a direct impact on the accuracy of the model - too many of them makes it harder for the model to choose the correct one. We therefore use the content of the database again in order to reduce candidates. In contrast to \emph{Value Candidate Generation} we do not use similarity metrics, but rather require exact matches. 

It is important to understand that we cannot validate all candidates in that way. Consider the following two examples: \emph{'List the top 3 albums of Elton John in the Billboard charts'} and \emph{'Find all albums of Elton John starting with "goodbye"'}. In these cases, we would not find \emph{'3'} or \emph{"goodbye"} in the database. In the first example, the value \emph{3} is not part of the database but is used in the SQL query to limit the results. In the second example the token \emph{"goodbye"} requires a wildcard match. Unfortunately, a wildcard match is not sufficient to validate a candidate, as it will provide too many false positives due to its flexibility. We therefore exclude numeric values and values extracted from quotes from the validation against the database.

During \emph{Value Candidate Validation} we also register in which table and column a value candidate is found.

\subsubsection{{\bf Value Candidate Encoding}}
All steps so far serve the purpose of establishing a solid set of value candidates as input to our neural architecture. It is after all the neural network that decides which value to choose. The pre-processing only fulfills the purpose of extracting and generating reasonable candidates.

The candidate encoding works similarly to the encoding of tables and columns. However, here we encode the location (i.e. table and column) where we found a candidate together with the value itself.

As an example consider the question \emph{"Show me all flights with destination Kennedy airport"} in Figure \ref{figure:value_candidates_encoding}. The value we are looking for is \emph{"JFK"}, which is contained in table \emph{Flight}, column \emph{Destination}. At the same time the term \emph{"Kennedy"} also appears in other tables, e.g. in table \emph{President}, column \emph{Name}.
Thanks to the additional table/column information, the encoder has the opportunity to build up attention (visualized by the black connectors) not only to the value itself, but also to the location where the value has been found. As this question contains the tokens \emph{"flights"} and \emph{"destination"}, the attention established with table \emph{Flight} is higher than to the other value candidate with table \emph{President} and column \emph{Name}.

Each value candidate together with its location, is separated from the other values by using the designated \emph{Separator} token of the encoder. Each value token is further tokenized in word pieces using the WordPiece\cite{WordPiece} segmentation algorithm. The input for the encoder is then a list of pre-trained embeddings, one for each word piece.

\subsubsection{Neural Architecture}
After encoding the value candidates we continue similar to \emph{ValueNet light} in Section \ref{subsec:ValueNetLight}. We use the neural architecture explained in Section \ref{subsec:NeuralArchitecture}.

\section{Experiments}
\label{sec:experiments}
In this section, we show the results of the experiments conducted with  \emph{ValueNet light} and \emph{ValueNet} for translating natural language questions to SQL. In particular, we will address the following research questions with respect to NL-to-SQL systems: (1) How well do our approaches \emph{ValueNet} and \emph{ValueNet light} perform  on the NL-to-SQL task incorporating values? (2) What is the difference in performance between \emph{ValueNet light} which starts with a list of values and \emph{ValueNet}, which must come up with a list of value candidates on its own? 

We will also show that both \emph{ValueNet light} and \emph{ValueNet} perform similarly or even better than state-of-the-art systems that are evaluated on the \emph{Execution Accuracy}.

\subsection{Dataset}
\label{subsec:Dataset}

For our experiments we use the Spider\cite{Spider} dataset which contains 10,181 natural language questions and their SQL equivalents. The queries have four levels of difficulty and contain most SQL operators (ORDER BY/GROUP BY/HAVING and nested queries). The queries are spread over 200 publicly available databases from 138 domains. Each database has multiple tables, with an average of 5.1 tables per database. The dataset is split into training set, validation (development) set and test set. The training set contains 8,659 queries, the validation set 1,034 queries. Note that we do not have access to the test set. The training set covers 146 databases while the validation set covers 20 different, i.e. unseen, databases. Hence, this dataset allows us to evaluate how well our two systems perform {\it transfer learning} between queries against databases in the training set and queries against {\it unseen databases} in the validation set. 

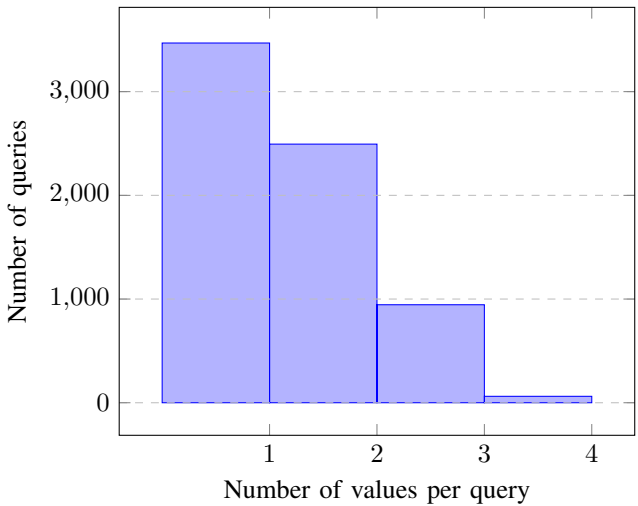
\begin{figure}[H]
\begin{tikzpicture}
\begin{axis}[
    area style,
    area style,
    xtick={1,2,3,4},
    ymajorgrids=true,
    grid style=dashed,
    xlabel={Number of values per query},
    ylabel={Number of queries},
    ]
\addplot+[ybar interval,mark=no] plot coordinates { 
    (0,3469)
    (1,2494)
    (2,945)
    (3,62)
    (4,30)
    };
\end{axis}
\end{tikzpicture}
\caption{\label{plot:value_distribution} Value distribution in the Spider data set. 3,469 samples contain no values. 2,494 samples contain one value, 945 two values, 62 three values and 30 samples contain 4 values.}
\end{figure}

We further analyzed the value distribution in the Spider dataset. We focused on the \emph{train} split, which contains exactly 7,000 NL questions\footnote{Although the training set of Spider already contains 7,000 novel questions, the authors of Spider added further 1,659 questions from existing datasets like e.g. IMDB adding up to a total of 8,659 question/query pairs to train on.} for the 10,181 samples. 
3,531 of the 7,000 sample questions contain values. Theses 3,531 sample questions contain a total of 4,690 values. See Figure \ref{plot:value_distribution} for a distribution of the values.

\subsubsection{Value Difficulty}
\label{subsubsec:ValueDifficulty}
The creators of the Spider dataset determined the difficulty level of a query without considering values. There is, however, a wide range of difficulty when it comes to extracting the correct values out of a question. We classify this difficulty into four levels:
\begin{itemize}
\item{\it Easy}: The value is clearly extractable by an NER system and is contained in the database as extracted. Example: \emph{"How many pets are owned by students that are older than 20?"} where the value is \emph{20}. 
\item{\it Medium}: The value is extractable by an NER system but might appear in a slightly different form in the database. Example: \emph{"What are the rooms for members of the faculty who are professors"} where the value in the database is \emph{Professor}. 
\item{\it Hard}: The value is extractable by an NER system but domain knowledge is needed to find the correct value. Example: \emph{"Show all flight numbers from Los Angeles."} where the value in the database is \emph{LAX}.
\item{\it Extra hard}: The value is not explicitly recognizable as value and therefore hard to extract. Example: \emph{"What are the names of nations where both English and French are official languages?"} where the values \emph{English} and \emph{French} can be directly extracted, but a third value in the database is 
\mintinline{SQL}{Language.IsOfficial = True}. 
\end{itemize}

%In Section \ref{subsec:Results} we will discuss  which difficulty levels can be handled by \emph{ValueNet}.

\subsubsection{Value Types}
We find the Spider dataset to contain a wide range of different values. The dataset includes but is not limited to: numeric values, strings and single characters, different ways of representing dates, times and duration, locations (e.g. addresses, countries, airports), specific codes (e.g. \emph{Airbus-A740} or \emph{CIS-220 }), status (e.g. \emph{successful} or \emph{completed}) and Boolean, names and salutations as well as e-mail addresses.  Despite some missing value types (e.g. phone numbers),  we consider the variability of value types to be comparable to a real-world environment.

\subsection{Evaluation Metrics}
\label{subsec:Metrics}
The Spider challenge\footnote{https://yale-lily.github.io/spider} comes with two different evaluation metrics: \emph{Exact Matching Accuracy} and \emph{Execution Accuracy}. As we briefly explained in Section \ref{sec:introduction}, \emph{Exact Matching Accuracy} compares the synthesized query to the gold query, while \emph{Execution Accuracy} requires executing the synthesized query against a database and compares if the result is the same as when executing the gold query.

To the best of our knowledge, we introduce the first NL-to-SQL system which is evaluated via \emph{Execution Accuracy} and whose source code is publicly available. Note that during the time of writing the paper, there are three systems that participated in the "Leaderboard - Execution with Values" with results from May 2020 (AuxNet + BART, BRIDGE + BERT and GAZP + BERT). However,as of now, these systems have not been published and the source code is not available, therefore, we are currently unable to reproduce their results. For this reason, we only compare our approaches against their reported accuracy in May 2020. 

\subsection{Experimental Setup}
\label{subsec:SetupMethods}

{\it Hardware}: All experiments were executed on a Tesla V100 GPU (32GB memory) with an Intel(R) Xeon(R) CPU E5-2650 v4 (4 cores) and 16GB memory.

{\it Frameworks}: The experiments are implemented using PyTorch. We use the code of IRNet\cite{IRNet} as the base for our implementation. For the encoder model (Section \ref{subsec:NeuralArchitecture}) we use the popular Transformer\cite{TransformerLibrary} library. For validation we use the official Spider validation script\footnote{https://github.com/taoyds/spider}.

{\it Implementation:} In our implementation we provide a transformer encoder which can be configured to use any modern pre-trained transformer model like RoBERTa\cite{RoBERTa} or XLNet\cite{XLNet}. To produce comparable results with state-of-the-art systems, we use the default \emph{BERT-Base} model for all experiments.

We further use bi-directional LSTM networks to summarize multi-token columns/tables/values (described in Section \ref{subsubsec:encoder}) with a dimensionality of 300. We use the same dimensionality for the decoder-network described in Section \ref{subsubsec:decoder}.

Moreover, we use an Adam\cite{ADAM} optimizer with three different learning rates: $2\mathrm{e}{-5}$ for the encoder, $1\mathrm{e}{-3}$ for the decoder and $1\mathrm{e}{-4}$ for the connection parameters in between. We further use a dropout of 0.3 and a batch size of 20. The learning rates for the encoder are the default parameters for BERT fine-tuning, all other hyperparameters have been set based on an empirical hyperparameter sweep.

To reproduce our experiments we release all code including hyperparameters on Github\footnote{https://github.com/brunnurs/valuenet}.

\subsection{Results}
\label{subsec:Results}

In this section we report the results of our two approaches \emph{ValueNet light} and \emph{ValueNet} on the Spider dataset using the \emph{Execution Accuracy} metric. This score includes (in contrast to the \emph{Exact Matching Accuracy} metric) the proper prediction of values.

As mentioned previously, there are currently no direct competitors using the \emph{Execution Accuracy} metric since neither the code nor the binaries are available to re-produce their experiments. However, as the Spider leaderboard {\it Execution with Values} though contains three candidates without publications ({\it GAZP + BERT, BRIDGE + BERT, AuxNet + BART}), we add these experiments as single data points.

\begin{figure}[H]
\begin{tikzpicture}
\begin{axis}[
    title={\textbf{\emph{ValueNet} vs. \emph{ValueNet light}}},
    xlabel={Number of epochs},
    ylabel={Accuracy},
    xmin=0, xmax=100,
    ymin=35, ymax=70,
    xtick={0,10,20,30,40,50,60,70, 80, 90},
    ytick={10,20,30,40,50,60,70},
    legend pos=south east,
    ymajorgrids=true,
    grid style=dashed,
]
\addplot[
    color=blue,
    ]
    coordinates {
    (1, 44.53875968992248)
    (2, 52.0968992248062)
    (3, 55.875968992248055)
    (4, 57.91085271317829)
    (5, 56.45736434108527)
    (6, 58.87984496124031)
    (7, 58.298449612403104)
    (8, 59.17054263565892)
    (9, 59.17054263565892)
    (10, 60.333333333333336)
    (11, 60.04263565891473)
    (12, 58.395348837209305)
    (13, 62.75581395348837)
    (14, 61.30232558139535)
    (15, 61.78682170542635)
    (16, 60.81782945736435)
    (17, 61.883720930232556)
    (18, 62.17441860465116)
    (19, 61.399224806201545)
    (20, 60.91472868217055)
    (21, 63.04651162790697)
    (22, 61.399224806201545)
    (23, 62.56201550387597)
    (24, 62.56201550387597)
    (25, 62.17441860465116)
    (26, 63.04651162790697)
    (27, 62.17441860465116)
    (28, 62.75581395348837)
    (29, 62.56201550387597)
    (30, 61.78682170542635)
    (31, 62.36821705426356)
    (32, 63.531007751937985)
    (33, 62.75581395348837)
    (34, 62.85271317829457)
    (35, 62.85271317829457)
    (36, 63.143410852713174)
    (37, 63.143410852713174)
    (38, 63.91860465116279)
    (39, 63.337209302325576)
    (40, 62.17441860465116)
    (41, 63.434108527131784)
    (42, 63.627906976744185)
    (43, 63.82170542635659)
    (44, 65.08139534883721)
    (45, 64.01550387596899)
    (46, 63.724806201550386)
    (47, 65.17829457364341)
    (48, 64.5)
    (49, 64.6937984496124)
    (50, 64.4031007751938)
    (51, 63.91860465116279)
    (52, 65.17829457364341)
    (53, 64.2093023255814)
    (54, 64.5)
    (55, 65.08139534883721)
    (56, 64.4031007751938)
    (57, 64.98449612403101)
    (58, 63.82170542635659)
    (59, 63.337209302325576)
    (60, 64.3062015503876)
    (61, 64.4031007751938)
    (62, 65.08139534883721)
    (63, 64.3062015503876)
    (64, 65.17829457364341)
    (65, 64.4031007751938)
    (66, 64.98449612403101)
    (67, 65.565891472868216)
    (68, 65.565891472868216)
    (69, 66.05038759689923)
    (70, 65.759689922480625)
    (71, 65.275193798449614)
    (72, 65.468992248062015)
    (73, 65.372093023255815)
    (74, 65.17829457364341)
    (75, 65.08139534883721)
    (76, 65.08139534883721)
    (77, 65.08139534883721)
    (78, 65.565891472868216)
    (79, 64.7906976744186)
    (80, 64.3062015503876)
    (81, 64.1124031007752)
    (82, 64.7906976744186)
    (83, 64.5)
    (84, 64.5)
    (85, 64.6937984496124)
    (86, 64.7906976744186)
    (87, 64.4031007751938)
    (88, 64.2093023255814)
    (89, 65.372093023255815)
    (90, 64.7906976744186)
    (91, 64.5968992248062)
    (92, 64.3062015503876)
    (93, 65.275193798449614)
    (94, 64.98449612403101)
    (95, 64.2093023255814)
    (96, 64.01550387596899)
    (97, 64.6937984496124)
    (98, 64.3062015503876)
    (99, 64.7906976744186)
    (100, 64.3062015503876)
    };

\addplot[
    color=red
    ]
    coordinates {
    (1, 42.69767441860465)
    (2, 50.54651162790697)
    (3, 51.51550387596899)
    (4, 54.32558139534884)
    (5, 54.616279069767444)
    (6, 59.84883720930233)
    (7, 55.58527131782945)
    (8, 56.166666666666664)
    (9, 58.298449612403104)
    (10, 58.298449612403104)
    (11, 57.13565891472868)
    (12, 57.62015503875969)
    (13, 58.87984496124031)
    (14, 57.03875968992248)
    (15, 58.201550387596896)
    (16, 60.624031007751945)
    (17, 58.201550387596896)
    (18, 57.32945736434108)
    (19, 60.52713178294574)
    (20, 59.46124031007752)
    (21, 57.52325581395349)
    (22, 58.87984496124031)
    (23, 59.17054263565892)
    (24, 59.94573643410853)
    (25, 61.399224806201545)
    (26, 59.751937984496124)
    (27, 59.84883720930233)
    (28, 61.30232558139535)
    (29, 59.46124031007752)
    (30, 60.624031007751945)
    (31, 59.94573643410853)
    (32, 59.17054263565892)
    (33, 60.333333333333336)
    (34, 59.84883720930233)
    (35, 58.97674418604651)
    (36, 58.87984496124031)
    (37, 60.43023255813954)
    (38, 61.30232558139535)
    (39, 61.10852713178295)
    (40, 62.46511627906976)
    (41, 61.20542635658915)
    (42, 59.46124031007752)
    (43, 61.01162790697675)
    (44, 60.720930232558146)
    (45, 61.01162790697675)
    (46, 61.496124031007746)
    (47, 61.78682170542635)
    (48, 61.59302325581395)
    (49, 61.496124031007746)
    (50, 62.56201550387597)
    (51, 60.52713178294574)
    (52, 60.720930232558146)
    (53, 61.496124031007746)
    (54, 61.68992248062015)
    (55, 62.46511627906976)
    (56, 61.68992248062015)
    (57, 61.01162790697675)
    (58, 60.81782945736435)
    (59, 61.68992248062015)
    (60, 60.333333333333336)
    (61, 62.17441860465116)
    (62, 61.30232558139535)
    (63, 61.59302325581395)
    (64, 61.496124031007746)
    (65, 61.20542635658915)
    (66, 60.91472868217055)
    (67, 62.36821705426356)
    (68, 62.17441860465116)
    (69, 61.10852713178295)
    (70, 61.78682170542635)
    (71, 61.59302325581395)
    (72, 61.30232558139535)
    (73, 62.27131782945736)
    (74, 61.98062015503876)
    (75, 60.720930232558146)
    (76, 61.78682170542635)
    (77, 61.59302325581395)
    (78, 61.399224806201545)
    (79, 61.399224806201545)
    (80, 61.496124031007746)
    (81, 61.399224806201545)
    (82, 61.30232558139535)
    (83, 61.78682170542635)
    (84, 61.496124031007746)
    (85, 61.59302325581395)
    (86, 61.59302325581395)
    (87, 60.91472868217055)
    (88, 61.10852713178295)
    (89, 61.01162790697675)
    (90, 61.399224806201545)
    (91, 61.68992248062015)
    (92, 61.68992248062015)
    (93, 61.59302325581395)
    (94, 61.496124031007746)
    (95, 62.07751937984496)
    (96, 61.399224806201545)
    (97, 61.68992248062015)
    (98, 62.56201550387597)
    (99, 61.98062015503876)
    (100, 61.68992248062015)
    };

\addplot[
    color=magenta,
    mark=otimes,
    ]
    coordinates {
    (51, 53.5)
    };
    
\addplot[
    color=green,
    mark=otimes,
    ]
    coordinates {
    (51, 59.9)
    };
    
\addplot[
    color=black,
    mark=otimes,
    ]
    coordinates {
    (51, 62.6)
    };

\legend{ValueNet light, ValueNet, GAZP + BERT, BRIDGE + BERT, AuxNet + BART}    
\end{axis}
\end{tikzpicture}
\caption{\label{plot:valueNetVsValueNetLight}The performance of \emph{ValueNet light} and \emph{ValueNet} on the Spider data set using the \emph{Execution Accuracy} evaluation metric. We represent our competitors on the Spider leaderboard with three single values due to their unpublished papers/code. We only visualize accuracy scores in the range of 35 to 70\% to emphasize the difference (see y-axis). The reported numbers are an average of five runs.}
\end{figure}
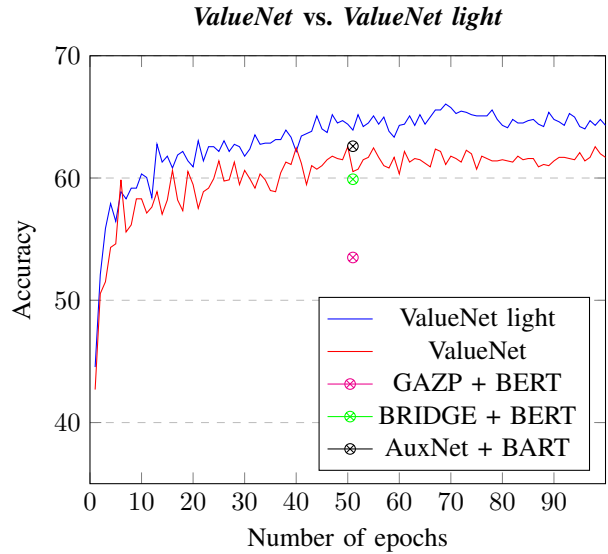

As we see in Figure \ref{plot:valueNetVsValueNetLight} both \emph{ValueNet} and \emph{ValueNet light} outperform {\it GAZP + BERT} and {\it BRIDGE + BERT}. The more advanced model {\it AuxNet + BART} levels on score with our ValueNet implementation. However, \emph{ValueNet light} also outperforms AuxNet + BART.

Note that we use the smallest version of BERT\cite{BERT} as encoder, whereas AuxNet + BART uses a much more advanced pretrained language model as encoder, namely BART \cite{BART}. We therefore expect our results to be even higher when augmenting the encodings with BART instead of BERT.

\subsection{ValueNet vs. ValueNet light}
\label{subsec:ValueNet_ValueNetLight}
As expected there is a performance gap between \emph{ValueNet} and \emph{ValueNet light} of 3\%-4\%. There are two possible reasons for this performance gap:
%\begin{description}

(1) {\it Non-extractable values}: While in \emph{ValueNet light} there is a list of all used values provided, \emph{ValueNet} needs to extract these values first from the question as described in Section \ref{sec:ValueNet}. 
Let us understand how many values we lose during that process and keep in mind that each value, that we cannot extract, will result in a failed sample. 
For the train split of the Spider dataset, which includes 3,531 samples containing values (a total of 4,690 values), we manage to extract all values for 3,200 samples. That means \emph{ValueNet} is capable of extracting around 90\% of all values. This share of extractable values stays constant for the validation dataset. 

Referring to the value difficulty of Section \ref{subsubsec:ValueDifficulty} we found that almost all of the remaining 10\% not found values belong to the difficulty classes \emph{Hard} and \emph{Extra Hard}. For instance, in the question \emph{"What are the full names of all left handed players?"} we failed to extract the value \emph{'L'} which would match the table/column \emph{players.hand}.

(2) {\it (Too) Many value candidates}: \emph{ValueNet light} is provided with a list of exact values for a sample query and then has to select each of them at the right time when synthesizing the query. If we revisit Figure \ref{plot:value_distribution}, we see the distribution of values for all queries in the dataset. We can observe that the maximum number of values a sample contains is 4. We also see that the majority of samples has only 1 or 2 values. 

%The number of value candidates \emph{ValueNet light} has to pick from is therefore rather low. \emph{ValueNet}, on the other hand, can due to its generative approach (see Section \ref{subsubsec:CandidateGeneration}) come up with any number of value candidates. The process of choosing the right value out of these candidates by the neural model (see Section \ref{subsubsec:decoder}) is therefore more error-prone. We visualize the difference between the number of actual values (\emph{ValueNet light}) of a sample and the number of value candidates generated by \emph{ValueNet} in Figure \ref{plot:difference_values_vs_value_candidates}. We see that for most samples \emph{ValueNet} comes up with the right number of value candidates and thus the difference is zero. However, for around a third of the examples the neural model of \emph{ValueNet} has to pick from a larger number of value candidates than the neural model of \emph{ValueNet light}. 
%\end{description}

\subsection{Results by Difficulty}
The Spider evaluation metric defines the difficulty
based on the number of SQL components, selections, and conditions, so that queries that contain
more SQL keywords (\mintinline{SQL}{GROUP BY, ORDER BY,
INTERSECT}, nested subqueries, column selections and aggregators, etc) are considered to be
harder. Spider defines 4 levels of difficulty: {\it Easy, Medium, Hard} and {\it Extra hard}. Many of the {\it Hard/Extra hard} queries contain multiple SQL keywords in combination with nested subqueries. We now want to analyze the accuracy of translating from NL to SQL with respect to the difficulty of the query.

\begin{table}[H]
\caption{Accuracy for different types of queries grouped by difficulty.}
\label{table:results_difficulty}
        \begin{tabular}{p{2.5cm}|p{4cm}}
        \textbf{Difficulty} & \textbf{Accuracy} \\
        \hline
        Easy & 0.77  \\
        Medium & 0.62 \\
        Hard & 0.57 \\
        Extra-Hard & 0.43 \\
        \end{tabular}
\end{table}

Table \ref{table:results_difficulty} shows the translation accuracy of \emph{ValueNet} grouped by query difficulty. We can see that for easy queries, \emph{ValueNet} achieves an average accuracy of 77\%. For extra-hard queries, the average accuracy drops to 43\%.

\subsection{Error Analysis}
We analyzed the 352 failed examples of ValueNet on the development set. For about 50\% of all errors (176 samples) we did a thorough manual analysis and found the following main causes of errors. Be aware that multiple error causes can appear per example.

{\it Column and Table Prediction}: In 50\% of all analyzed errors ValueNet fails to predict the correct column. In around 25\% it chooses a column from another table, therefore the table prediction is also incorrect. The main reason for those errors is that columns across different tables have similar or even identical names and are thus hard to distinguish. Examples for such column names are \emph{name}, \emph{id} or \emph{description}, which appear usually in multiple tables. ValueNet, similar to IRNET\cite{IRNet}, struggles with such cases. Incorporating a more sophisticated schema linking approach, as for example proposed by RAT-Sql \cite{RatSQL}, might help to prevent such errors.

{\it Errors in the SQL Sketch}: In 39\% of all analyzed cases we find errors in the SQL sketch, i.e. the logical form of the query. It is important to note though that the majority (76\%) of these errors appear in queries classified as \emph{Hard} or \emph{Extra Hard} by the Spider authors. It is further interesting to see that we did not find a completely incorrect SQL sketch in any of the analyzed examples. We frequently found that the SQL Sketch was 80\% correct but included a minor mistake.

A clear pattern is hard to establish. Some of the \emph{Hard} and \emph{Extra Hard} examples require advanced common knowledge which is hard to incorporate into a model. However, some of the failed examples on lower difficulties could easily be solved with domain-specific training (e.g. "oldest player" is incorrectly interpreted as \mintinline{SQL}{ORDER BY birthdate ASC} rather than \mintinline{SQL}{DESC}).      

{\it Value Selection}: In 9\% of all analyzed errors ValueNet selects the wrong value. Note that a third of these cases leads back to one single value -- a company name called \emph{"JetBlue Airways"}. We assume that domain-specific fine tuning of the encoder on the database content could avoid such errors.

{\it False Negatives}: 9\% of all reported errors are false negatives. These examples range from missing or wrong data in the provided databases to mismatches between the question and the ground truth query. One common mistake is e.g. a missing table in the query, even though it is clearly stated in the question. 

In many cases, it is debatable if a question really gives a hint for a specific SQL clause or not. One example is the keyword \mintinline{SQL}{DISTINCT}, which is often hard to derive from a question if not specifically hinted. In this case, a more advanced error metric might be able to classify the degree of error. 

%\newpage
%\subsection{Interpretation of the Results}
%{\bf TODO: Do we still need these conclusions and comparisons to IRNet? We could drop these paragraphs without loosing any insights.}
%A first interesting insight is the relatively small difference between \emph{ValueNet light}, \emph{ValueNet}. The accuracies of our approaches lie in around 60\%, in an interval of 7\%. This is therefore surprising since we know that the share of samples containing values is around 50\%, so in half of the cases \emph{ValueNet light}/\emph{ValueNet} have to solve a quite different task from e.g. \emph{IRNet}. We therefore conclude that incorporating values is possible without reducing the prediction power of the model.

%A second insight is the relatively small difference (4\%) between \emph{ValueNet light} and \emph{ValueNet}. The conclusion here is that if we manage to come up with the correct value candidates (even if we propose some value candidates which are incorrect) the neural model is capable of selecting the correct one with a high probability. 

\subsection{End-to-End Performance: Translation Time}
An important question to answer is how long it takes {\it ValueNet} to synthesize a query given an question. At inference time this process happens interactively while the user waits for a result. The process therefore needs to be performant.

\begin{table}[h]
\caption{Translation Time}
\label{table:translation_time}
        \begin{tabular}{c|c|c}
        \textbf{Step} & \textbf{Average Time [ms]} & \textbf{Standard Deviation
 [ms]} \\
        \hline
        Pre-Processing & 80 & 5  \\
        Value lookup & 234 & 43 \\
        Encoder/Decoder& 76 & 14 \\
        Post-Processing & 13 & 2 \\
        Query-Execution & 15 & 3 \\
        \end{tabular}
\end{table}

In Table \ref{table:translation_time} we see the total translation time, split up by the process steps defined in Figure \ref{figure:architecture_overview}. The duration was measured over the execution of all 1,034 samples of the validation set. The overall translation time per query is on average 418 ms and therefore sufficiently fast for interactive usage. For use in larger databases, the process can further be improved by using an advanced inverted index when looking up values, as this step consumes by far the largest amount of time.

\section{Related Work}
\subsection{NL-to-SQL}
Since the end of the 1970s, building natural language interfaces for databases (NLIDBs) has been a significant challenge. Many of the early work \cite{EarlyWork1, EarlyWork2, EarlyWork3} focused on rule-based approaches with handcrafted features. Later systems enabled users to query the databases with {\it simple keywords} \cite{simitsis2008precis, blunschi2012soda, bast2015more}. The next step in the development was to enable processing more complex natural language questions by applying a {\it pattern-based approach} \cite{damljanovic2010natural, zheng2017natural}. Moreover, {\it grammar-based} approaches were introduced to improve the precision of natural language interfaces by restricting users to formulate queries according to certain pre-defined rules\cite{song2015tr, ferre2017sparklis}. Finally, a comprehensive overview of NLIDB systems is given in \cite{affolter2019comparative}.

While most of these approaches work well when customized for a specific database (with a restricted set of keywords or natural language patterns), they are often not competitive in a cross-domain setting with complex questions. One of the most advanced systems is currently Athena++\cite{sen2020athena++}. However, neither the source code nor the binaries of that system are publicly available for reproducing the results.

More recent approaches use advanced neural network architectures to synthesize SQL queries given a user question. The work of \cite{DongLapata} uses a classical encoder-decoder architecture based on Long Short Term Memory (LSTM) networks \cite{hochreiter1997long}. \emph{Seq2SQL}\cite{WikiSQL} adds a reinforcement learning approach to learn query-generation policies. That system creates a reward signal by executing queries against the database in-the-loop. \emph{SQLova}\cite{SQLova} is the first work to use a transformer-based encoder \cite{AttentionIsAllYouNeed} to solve the WikiSQL\cite{WikiSQL} challenge.

The Spider\cite{Spider} dataset, which covers 200 databases and more than 10,0000 training data samples, is currently considered the de-facto standard for evaluating NLIDBs (or NL-to-SQL-systems) based on machine learning approaches. A recent approach \cite{deriu2020methodology} introduces a novel framework for generating training data by inverting the data annotation process. The advantage of this approach is to generate training data more quickly and to cover a wider range of queries.

Let us now focus on recent systems that use the Spider dataset for evaluating the accuracy of generating SQL given a natural language question. IRNet\cite{IRNet}, for instance, uses a transformer encoder and a decoder based on an LSTM network. It further introduces an intermediate representation based on an abstract syntax tree as an alternative to directly synthesizing SQL. The main goal of this intermediate representation is the abstraction of SQL-specific implementation details. In our work we use and extend this approach to handle natural language queries that incorporate values, i.e. that require analyzing the base data of the database.

\emph{RAT-SQL}\cite{RatSQL} is another NL-to-SQL system that achieved state-of-the-art results on the Spider challenge. It focuses on the problem of \emph{schema encoding} and \emph{schema linking}. The work proposes a new relation-aware self-attention mechanism based on transformers with promising results on non-trivial database schemas. 

A good overview about the performance of the above-mentioned systems is given in \cite{kim2020natural}. The paper conducts a detailed experimental evaluation of both traditional and neural network-based systems whose source code or binaries are available for reproducibility studies. The paper concludes that there is still significant potential for improving current state-of-the-art systems to work in real-world environments.

%\subsection{Transformer Architectures}

%Based on the well-known paper "Attention is all you need" \cite{AttentionIsAllYouNeed}, transformers - a special type of neural network - started in mid 2017 to become one of the most promising techniques for natural language processing (NLP). One of the earliest systems was BERT \cite{BERT} which combined the transformer architecture with massive unsupervised pre-training. BERT was the first approach to achieve state-of-the-art results in a large number of NLP tasks. Succeeding works by \cite{XLNet, RoBERTa} achieved even higher results on NLP benchmarks, though mainly by using more training data, larger neural network architectures and novel pre-training approaches.

%While transformers are undeniably among the most significant achievements in the recent NLP landscape, they have also started a controversy about the \emph{more data/larger models/more computational power} mentality. Therefore, in addition to increasing model sizes like the authors of \cite{MegatronLM} did, research started also to develop leaner transformer models, which can be used on mobile devices or non-GPU servers at inference time \cite{DistilBERT, ALBERT}. 

%Transformer architectures have been used for different tasks such as natural language translation~\cite{AttentionIsAllYouNeed}, natural language generation~\cite{GPT} and recently also for entity matching as part of data integration~\cite{brunner2020entity}.

\subsection{Finding Matching Database Values}
The importance of values and the idea of finding correct values through database lookups was already known in works based on the WikiSQL challenge as the meta paper ~\cite{WhatItTakesToAchieve100WikiSQL} shows. With the Spider~\cite{Spider} challenge providing values in around 50\% of its training data samples, it is an ideal dataset to continue working on the challenge of building a real world NL-to-SQL system incorporating values. Unfortunately most works \cite{IRNet, RatSQL, GraphNeuralNetworks} on the Spider challenge chose an evaluation metric that does not consider values. Hence, with our two approaches presented in this paper, we deliver an end-to-end NL-to-SQL system incorporating values and hope to motivate further work to solve this challenge. Moreover, we provide the source code of our system such that the results can be reproduced by other researchers.
\section{Conclusions \& Future Work}
In this work we propose \emph{ValueNet light} and \emph{ValueNet} -- two end-to-end NL-to-SQL systems incorporating values. We evaluate them on the Spider dataset and demonstrate that incorporating values does not affect the accuracy of translating from natural language to SQL negatively. We achieve state-of-the-art results and provide, to our knowledge, the first detailed discussion and source code for an NL-to-SQL system that {\it synthesizes a full SQL query including values} on
the challenging Spider dataset

In particular, with \emph{ValueNet} we propose an architecture sketch to come up with good value candidates. These value candidates are then incorporated into the end-to-end query translation process through the neural model. Generating good value candidates is difficult as the values extracted from a question often differ from the actual values in the database. We thus use the database content, in combination with a generative approach, to produce promising value candidates. Finally, the neural network decides which of these value candidates is the best match for the intention of the natural language question from the user.

Moreover, \emph{ValueNet} is a system that synthesizes complete queries which can be executed against a database. In contrast to many recent works, we provide an approach which can be used in a real-world scenario. We hope to motivate further papers on solving this challenge to use the Spider \emph{Execution Accuracy} metric.

As part of future work, we will further improve the architecture sketch of how to come up with good value candidates. One possible avenue of research  is to apply a generative neural network approach (e.g. based on text GANs\cite{GANs}) in combination with the available data from the database.

% ensure same length columns on last page (might need two sub-sequent latex runs)
% \balance

%ACKNOWLEDGMENTS are optional
\section*{Acknowledgments}
This project has received funding from the European Union’s Horizon 2020 research and innovation program under grant agreement No 863410. We thank Kate Kosten for linguistic improvements of the paper as well as Ana Sima and Georgia Koutrika for fruitful discussions.

\bibliographystyle{abbrv}
\bibliography{0_main}

%\begin{thebibliography}{00}
%\bibitem{b1} G. Eason, B. Noble, and I. N. Sneddon, ``On certain integrals of Lipschitz-Hankel type involving products of Bessel functions,'' Phil. Trans. Roy. Soc. London, vol. A247, pp. 529--551, April 1955.
%\end{thebibliography}
%\vspace{12pt}

\end{document}